\documentclass[12pt]{article}
\usepackage{epsf}
\usepackage{amsfonts} 
\usepackage{amsbsy} 

\newdimen\hsgraph \newdimen\vsgraph
\hsgraph=\hsize \advance\hsgraph by-1.2truein
\vsgraph=\vsize \advance\vsgraph by-1.5truein

\newcommand{\C}[1]{{\mathcal #1}}

\newcommand{\R}[1]{{\mathrm #1}}

\newcommand{\beq}{\begin{equation}}
\newcommand{\eeq}{\end{equation}}
\newcommand{\bea}{\begin{eqnarray}}
\newcommand{\eea}{\end{eqnarray}}
\newcommand{\rf}[1]{(\ref{#1})} 

\newcommand{\mod}[1]{{\vert #1\vert}}

\newcommand{\mubar}{{\bar\mu}}

\newcommand{\nn}{\nonumber}
\newcommand{\Hbar}{{\widetilde{\C H}}}

\begin{document}
\topmargin 0pt
\oddsidemargin 5mm
\headheight 0pt
\topskip 0mm

\addtolength{\baselineskip}{0.20\baselineskip}

\pagestyle{empty}

\begin{flushright}
OUTP-98-78P\\
16th November 1998\\
hep-th/9811205
\end{flushright}

\begin{center}

\vspace{18pt}
{\Large \bf The Hausdorff dimension in polymerized quantum gravity}

\vspace{2 truecm}

{\sc Martin G. Harris\footnote{e-mail: harris@thphys.ox.ac.uk} 
and John F. Wheater\footnote{e-mail: j.wheater1@physics.ox.ac.uk}}

\vspace{1 truecm}

{\em Department of Physics, University of Oxford \\
Theoretical Physics,\\
1 Keble Road,\\
 Oxford OX1 3NP, UK\\}

\vspace{3 truecm}

\end{center}

\noindent
{\bf Abstract.} We calculate the Hausdorff dimension, $d_H$, and the 
correlation function exponent, $\eta$, for polymerized two dimensional
quantum gravity models.  If the non-polymerized model has correlation function exponent $\eta_0 >3$ then $d_H=\gamma^{-1}$ where $\gamma$ is the susceptibility exponent. This suggests that these models may be in the same universality
class as certain non-generic branched polymer models.

\vfill
\begin{flushleft}
\end{flushleft}
\newpage
\setcounter{page}{1}
\pagestyle{plain}

The idea of polymerization was first introduced in the context of matrix
models \cite{das,gaume,korchemsky} and then later generalized to an arbitrary
random surface theory (ie model of discretized two dimensional quantum gravity)
by Durhuus \cite{durhuus}.  The basic idea is that the random surface 
(universe) can have other random surfaces (baby universes) attached with the
minimal possible contact (this is defined more carefully below). Depending
upon the fugacity for these contact terms the system can exist in one of
three phases; at low fugacity there is a single random surface with few outgrowths attached, at high fugacity there are many outgrowths and the system is 
a generic branched polymer, and at an intermediate point there is a branched polymer
structure in which each node is itself a critical surface. This  point
is the polymerized model and has been studied in a number of papers \cite{adj,
klebanov,tjjfw,Harris}. The polymerized models have the interesting feature that
their susceptibility exponent is positive but less than the generic branched
polymer value $1/2$. In this letter we compute the Hausdorff dimension
of these polymerized models. 

In general we define the grand canonical partition function for an 
ensemble of graphs $\C G$ by
\beq \C Z(\mu)=\sum_{\R G \in \C G} e^{-\mu\mod {\R G}}w_{\R G}\label{1}\eeq
where $\mod {\R G}$ denotes the number of points in G, and $w_{\R G}$ the
weight for the graph G 
 (for an introduction to this material see for example \cite{book}).  We will assume throughout that graphs are defined with
one marked point. The sum in \rf{1} is convergent for $\mu>\mu_c$ and as
 $\mu\to\mu_c$ we expect
\beq \C Z(\mu)\simeq R_1-R_2(\mu-\mu_c)^{1-\gamma}\label{1a}\eeq
where $R_{1,2}$ are regular functions of $\mu$. In addition we define
the susceptibility
\beq \chi(\mu)=-\frac{\partial \C Z(\mu)}{\partial \mu}\sim A-B(\mu-\mu_c)^{-\gamma}\label{1b}\eeq
where $A$ and $B$ are constants.

Now consider an ensemble of graphs $\C G$ constructed from a base graph
ensemble $\C G_0$ and a ``baby'' graph ensemble  $\C G_1$; all the graphs
in  $\C G_{0,1}$ are assumed to have weight $w_{\R G}=1$.
As shown in fig.1, baby graphs are attached to the base
 graph with a fugacity $\rho$
 by  identifying their marked point with a point in the base graph
(this is slightly different from the usual construction but more convenient
for our purpose). For a given base graph  points such as `a', to which no
baby is attached, contribute a factor 1 to $w_{\R G}$, whereas points
`b'
contribute a factor $\rho e^\mu \C Z_1(\mu)$ once all baby graphs have
been summed over. The extra factor of $ e^\mu$ multiplying $ Z_1(\mu)$  
 arises
because we are identifying points in the base and baby graphs. This reduces
the total number  of points in the product graph by one in comparison to the 
base and baby taken separately. Summing over all possible ways of attaching
the babies to the base graph we obtain
\bea \C Z(\mu)&=&\sum_{\R G \in \C G_0} e^{-\mu\mod {\R G}}
\left(1+\rho e^\mu \C Z_1(\mu)\right)^{\mod {\R G}}\cr
&=&\C Z_0(\mubar)\label{2}\eea
where 
\beq \mubar=\mu-\log\left(1+\rho e^\mu \C Z_1(\mu)\right).\label{3}\eeq
The structure defined
in \rf{2} and \rf{3} is identical to that derived in \cite{durhuus}.
Throughout this letter we will denote the value of $\mubar$ at which
$\C Z_0(\mubar)$ is non-analytic by $\mubar_0$.

\begin{figure}[t]

\hfil{\epsfxsize=5cm \epsfbox{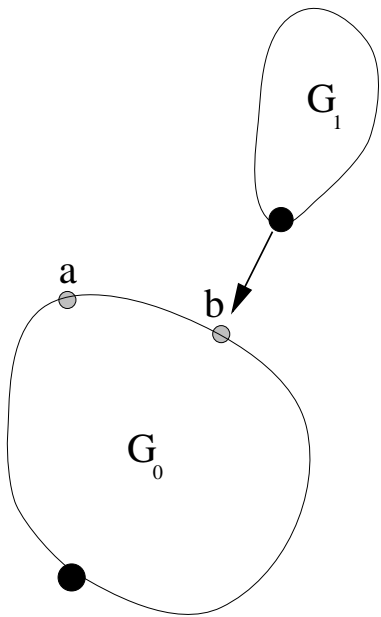}}\hfil
\caption{Attaching a baby ${\R G}_1\in{\C G}_1$ to the base graph ${\R G}_0\in{\C G}_0$. The solid circles denote
the marked points on the two graphs and the shaded circles points on the 
base graph.}
\end{figure}

Now consider the two-point function
\beq \C H(R,\mu)=\sum_{\R G \in \C G} e^{-\mu\mod {\R G}}\sum_{i\in \R G}
\delta(d_{\R G}(i)-R)\label{4}\eeq
where $d_{\R G}(i)$ is the geodesic distance of the point $i$ from the marked
point on G. Note that 
\beq \sum_{R=0}^\infty\C H(R,\mu)=-\frac{\partial \C Z(\mu)}{\partial \mu}=\chi(\mu).\label{6a}
\eeq
We expect that $\C H$ has the asymptotic behaviour \cite{AmbA,AmbE}
\bea \C H(R,\mu)\sim& e^{-m(\mu) R},&\quad m(\mu) R>>1 \nn\\
\sim& R^{1-\eta},&\quad  m(\mu)^{-1} >> R>>1.\label{4a}\eea
 As $\mu\to\mu_c$, the mass gap vanishes as
\beq m(\mu)\sim (\mu-\mu_c)^\nu\label{4b}\eeq
where the correlation length exponent is related to the Hausdorff dimension
$d_H$ by $d_H\nu=1$. $\C H$ is related by discrete
 Laplace transformation to the
canonical ensemble quantity $n(R,N)$; this is the expectation on graphs
with a fixed number of vertices, $N$, of the number of vertices a geodesic 
distance $R$ from the fixed point. It is expected that 
\beq n(R,N)\sim R^{d_h-1}f\left(\frac{R^{d_H}}{N}\right)\label{4c}\eeq
where $f(x)$ is $O(1)$ at small $x$ and vanishes at large $x$ \cite{AmbE}.
The Hausdorff dimension $d_h$ is not necessarily the same as $d_H$
(for example on multi-critical branched polymers they are different
 \cite{BP2,AmbC}).

If G is constructed from a base graph with babies there are two
sorts of contribution to \rf{4}. The first is where the point
 $i$ lies in the base graph.
The second is where the point $i$ lies in a baby $\R G_1$
 which is attached to the base graph $\R G_0$  at $k$ say; in this case we have
\beq d_{\R G}(i)=d_{\R G_0}(k)+d_{\R G_1}(i)\label{5}\eeq
and of course G must contain at least one baby.
Attaching babies to points in the base graph as described above
(fig.1)
we then find that 
\bea\C H(R,\mu)&=&\sum_{\R G \in \C G_0} e^{-\mu\mod {\R G}}
\left(1+\rho e^\mu \C Z_1(\mu)\right)^{\mod {\R G}}\sum_{i\in \R G}\delta(d_{\R G}(i)-R)\nn\\
&&+\sum_{r=1}^R\sum_{\R G \in \C G_0} e^{-\mu\mod {\R G}}
\left(1+\rho e^\mu \C Z_1(\mu)\right)^{\mod {\R G}-1}
\sum_{k\in \R G_0}\delta(d_{\R G}(k)-R+r)\nn\\
&&\qquad\times\rho e^\mu\sum_{\R G' \in \C G_1} e^{-\mu\mod {\R G'}}\sum_{i\in \R G'}
\delta(d_{\R G'}(i)-r)\nn\\
&=&\C H_0(R,\mubar)+\sum_{r=1}^R\frac{\C H_0(R-r,\mubar)
\rho e^\mu\C H_1(r,\mu)}{1+\rho e^\mu \C Z_1(\mu)}.\label{6}\eea
%
It is straightforward to check that this expression satisfies \rf{6a}.

From the point of view of quantum gravity we are interested in the case where the ensemble $\C G_1$  from which the babies are drawn is the same as the full ensemble $\C G$; this means that we set $\C Z_1(\mu)=\C Z(\mu)$, and
$\C H_1(r,\mu)=\C H(r,\mu)$ in \rf{2}, \rf{3} and \rf{6}. Together with the 
assumption that  the susceptibility exponent of the $\C G_0$ ensemble satisfies
$\gamma_0<0$, this defines the polymerized quantum gravity models \cite{durhuus}.  We then obtain 
from \rf{3} the useful result that 
\beq 1+\rho e^\mu \C Z(\mu)= \left(1-\rho e^\mubar \C Z_0(\mubar)\right)^{-1}.
\label{7a}\eeq
To analyze the 
critical exponents of the polymerized theory we start by recalling the calculation of $\gamma$ \cite{durhuus}. By differentiating \rf{2} we obtain
\beq \chi(\mu)=\frac{\chi_0(\mubar)\left(1-\rho e^\mubar \C Z_0(\mubar)\right)}
{1-\rho e^\mubar \chi_0(\mubar)}.\label{7}\eeq
We can identify three regimes.\\
i) When $\rho$ is small enough the first singularity encountered on the
right hand side of \rf{7} as $\mu$ decreases is in $\chi_0(\mubar)$
at $\mubar=\mubar_0$
(note that since $\gamma_0<0$, $\chi_0(\mubar_0)$ is finite). From \rf{3} we have that as $\mu\to\mu_c$
\beq \mubar-\mubar_c=\mu-\mu_c-\log\left(1-C (\mu-\mu_c)^{(1-\gamma)}\right)
\label{9}\eeq
where $C$ is a constant.
Because $ \chi(\mu)$ is not divergent, $\gamma<0$ and therefore
$\mubar-\mubar_c\sim\mu-\mu_c$ and hence, using \rf{1b}, we have
 $\gamma=\gamma_0$.\\
ii) When $\rho$ is very large the denominator in \rf{7} will vanish 
at a  value of $\mu$ such that $\mubar>\mubar_0$, and $\chi_0(\mubar)$
is therefore still analytic, so we obtain
\beq  \chi(\mu)\sim\frac{1}{\mubar-\mubar_c}.\label{9a}\eeq
Now $ \chi(\mu)$ is  divergent so  $\gamma>0$ and hence from \rf{9} we obtain
\beq \mubar-\mubar_c\sim (\mu-\mu_c)^{(1-\gamma)}\label{10}\eeq
 so that 
\beq  \chi(\mu)\sim\frac{1}{(\mu-\mu_c)^{(1-\gamma)}}\label{9b}\eeq
and hence, using the  definition \rf{1b},
we find that $\gamma=1/2$ which is the standard branched polymer 
exponent.\\
iii) At some critical value $\rho=\rho_c$ the susceptibility $\chi_0(\mubar)$
will be non-analytic \emph{and} the denominator in \rf{7} will vanish at
the same value of $\mu$. In this case we find
\beq  \chi(\mu)\sim\frac{1}{(\mubar-\mubar_c)^{-\gamma_0}}.
\label{8}\eeq
Again using \rf{10} we get 
\beq  \chi(\mu)\sim\frac{1}{(\mu-\mu_c)^{-\gamma_0(1-\gamma)}}
\label{11}\eeq
and hence
\beq \gamma=\frac{-\gamma_0}{1-\gamma_0}.\label{12}\eeq
This result applies provided $\gamma_0>-1$. If $\gamma_0<-1$
then the denominator in \rf{8} is linear in $\mubar-\mubar_c$
and we obtain $\gamma=1/2$.

Now we can study $\C H(R,\mu)$ in these three regimes. Since \rf{6}
is a convolution it is convenient to introduce
\beq \Hbar(\omega,\mu)=\sum_{r=0}^\infty \C H(r,\mu)e^{-\omega r}\label{13}\eeq
and similarly for $\C H_0(r,\mu)$. Then from \rf{6} we obtain
\beq \Hbar(\omega,\mu)=\frac{\left(1-\rho e^\mubar \C Z_0(\mubar)\right)
\Hbar_0(\omega,\mubar)}{1-\rho e^\mubar \Hbar_0(\omega,\mubar)}\label{14}\eeq
which reproduces \rf{7} if we set $\omega=0$. From the general behaviour
\rf{4a}  we expect  that  
the singularities in $\omega$ of $\Hbar(\omega,\mu)$ will lie on the
negative real axis. The dominant physical behaviour is given by the nearest
singularity to the origin (or more precisely by the singularities that
converge onto the origin as $\mu\to\mu_c$).
To establish $d_H$ it
 is sufficient
to establish the behaviour of the location of this singularity,
 $\omega_0$, 
  because by \rf{4a} we expect $\omega_0=-m\sim -(\mu-\mu_c)^\nu$.

In regime i) $\Hbar_0(\omega,\mubar)$ has a singularity at
 $\omega\sim -(\mubar-\mubar_0)^{\nu_0}$.
However, the denominator of \rf{14} takes the same value at $\omega=0$
as does the denominator of \rf{7} and is therefore finite and positive; thus
$\Hbar(\omega,\mu)$ is singular at the same value of $\omega$ as
$\Hbar_0(\omega,\mubar)$ and  $d_H=d_{H_0}$.
The $\eta$ exponent is found by considering \rf{14} at $\mu=\mu_c$; from
\rf{4a} we get
\bea \Hbar_0(\omega,\mubar_c) &\sim& \sum_{r=1}^\infty r^{1-\eta_0}
 e^{-\omega r}\nn\\
&=&\Hbar_0(0,\mubar_c)-b\omega+\ldots +c\omega^{\eta_0-2}+\ldots,\label{X}\eea
where $b$ and $c$ are constants, and a similar expression for $\Hbar(\omega,\mu_c)$; we see that $\eta$ controls the leading non-analyticity in $\omega$.
 Substituting \rf{X} in \rf{14} we see that  $\eta=\eta_0$.
 We might also conclude that $d_h=d_{h_0}$ on the grounds that 
 $\Hbar(\omega,\mu)$ has the same leading
singularity structure as $\Hbar_0(\omega,\mubar)$ but with a multiplicatively
modified residue; however this is not a complete proof since the inverse
Laplace transform required to extract $n(R,N)$ from $\Hbar(\omega,\mu)$ is rather delicate.

In regime ii) we know that the denominator in \rf{14} vanishes at $\omega=0$
when $\mubar >\mubar_0$ and $\Hbar_0(\omega,\mubar)$ is analytic in the 
vicinity of $\omega=0$. Thus we can write
\beq \left(\Hbar_0(\omega,\mubar)\right)^{-1}=\left(\Hbar_0(0,\mubar)\right)^{-1}\left(1+a\omega+\dots\right)\label{15}\eeq
where $a$ is finite and the series is convergent; substituting this in \rf{14}
we get
\beq \Hbar(\omega,\mu)=\frac{\left(1-\rho e^\mubar \C Z_0(\mubar)\right)
\Hbar_0(0,\mubar)}{1+a\omega-\rho e^\mubar \Hbar_0(0,\mubar)}\label{16}\eeq
so there is a simple pole at 
\beq \omega\sim-(\mubar-\mubar_c)\sim-\sqrt{\mu-\mu_c}.\label{17}
\eeq
$\Hbar(\omega,\mu)$ thus consists of a simple pole plus a regular
function of $\omega$.  The long distance behaviour is determined
entirely by the simple pole 
so we have  pure generic branched polymer behaviour for which 
 $d_H=2$, $d_h=2$, and $\eta=1$ \cite{BP2}.

At $\rho=\rho_c$, in regime iii), the singularity in $\Hbar_0(\omega,\mubar)$ and the zero
of the denominator in \rf{14} coincide when $\mu=\mu_c$.
 The location of
the zero  is given by
\beq \Hbar_0(\omega_0,\mubar)=\frac{1}{\rho e^\mubar}.\label{22}\eeq
Now we exploit the fact that  $\Hbar_0(\omega,\mubar)$ is an analytic function of $\omega$ when $\mod\omega << (\mubar-\mubar_c)^{\nu_0}$ (this is because
its first singularity occurs at $\omega\sim -(\mubar-\mubar_c)^{\nu_0}$).
So, assuming that $\omega_0$ lies in the region where $\Hbar_0(\omega,\mubar)$
is analytic, and  expanding \rf{22} we obtain
\beq \Hbar_0(0,\mubar)+\sum_{n=1}^\infty
\frac{\omega_0^n}{n!}
\frac{\partial^n \Hbar_0(\omega,\mubar)}{\partial\omega^n}\Bigg |_{\omega=0}
=\frac{1}{\rho e^\mubar}.\label{23}\eeq
 From the definition \rf{13} we have that
\beq \frac{\partial^n \Hbar_0(\omega,\mubar)}{\partial\omega^n}\Bigg
 |_{\omega=0}=\sum_{r=0}^\infty (-r)^n \C H_0(r,\mubar).\label{24}\eeq
We can use  the asymptotic behaviour \rf{4a} to  deduce the leading
behaviour of these moments of $\C H_0(r,\mubar)$. For example we can compute
the moments of the trial function
\bea \C H_{trial}(R,\mu)=& m(\mu)^{\eta-1} e^{1-m(\mu) R},
&\quad m(\mu) R\ge1 \nn\\
=& R^{1-\eta},&\quad  m(\mu)^{-1} > R\ge 1\label{T}\eea
 which has the correct asymptotic behaviour
and is continuous; it is easy to check that, assuming Fisher scaling, the
zeroth moment has the same form as the susceptibility. For the higher
moments we  find
\bea \frac{\partial^n \Hbar_0(\omega,\mubar)}{\partial\omega^n}\Bigg
 |_{\omega=0}\sim& a_n,\hfil &\quad n<\eta_0-2\nn\\
\sim& a_n\log(\mubar-\mubar_c),&\quad n=\eta_0-2\nn\\
\sim& a_n (\mubar-\mubar_c)^{-\gamma_0-n\nu_0},&\quad n>\eta_0-2\label{30a}\eea
where  the $a_n$ are constants.
In the pure gravity case $\eta_0=4$ and so \rf{23} gives
\beq \omega_0 a_1+\omega_0^2 a_2\log(\mubar-\mubar_c) +\omega_0^3a_3(\mubar-\mubar_c)^
{-\gamma_0-3\nu_0}+\ldots\sim- (\mubar-\mubar_c)^{-\gamma_0}\label{31}\eeq
and hence
\beq \omega_0\sim - (\mubar-\mubar_c)^{-\gamma_0}.\label{25}\eeq
Note that \rf{25} is consistent with the assumption that 
$\mod{\omega_0} << (\mubar-\mubar_c)^{\nu_0}$. It is easy to check that 
provided $\eta_0>3$ this procedure is consistent and that  \rf{25} is the
correct solution. It follows immediately that
\beq \nu=-\frac{\gamma_0}{1-\gamma_0}=\gamma\label{32}\eeq
and therefore that 
\beq d_H=\frac{1}{\gamma}.\eeq
If $\eta_0\le 3$ then $\nu_0\ge-\gamma_0$ and hence the dominant singularities
in \rf{23} must occur at $\omega_0\sim -(\mubar-\mubar_c)^{\nu_0}$ and so
\beq d_H=\frac{1-\gamma_0}{\nu_0}.\eeq

As usual the $\eta$ exponent is found by considering \rf{14} at $\mu=\mu_c$.
Using \rf{X} we see that 
if $\eta_0>3$ the  term linear in $\omega$ dominates as $\omega\to 0$ and,
substituting into \rf{14}, we get
\beq \Hbar(\omega,\mu_c)\sim\frac{1}{\omega}\eeq
and hence that $\eta=1$; thus the Fisher scaling relation $\nu(2-\eta)=\gamma$
is satisfied. On the other hand if  $\eta_0\le 3$ then the non-analytic term
in \rf{X} dominates and we get
\beq \Hbar(\omega,\mu_c)\sim\frac{1}{\omega^{\eta_0-2}}\eeq
and hence that $\eta=4-\eta_0$; again it is easy to check that Fisher scaling
is obeyed.

There are at present only two cases for which we know the Hausdorff
dimension of the base graph. The first is that of pure (ie $c=0$) 
quantum gravity \cite{Kawai,AmbA} for which $d_{H_0}=4$, $\eta_0=4$, $\gamma_0=-1/2$;
at $\rho=\rho_c$ the polymerized model has $\gamma=1/3$ \cite{durhuus,adj} and, by our results,
$d_H=3$ and $\eta=1$. In fact 
these results for a polymerized system 
with pure gravity base graphs have been obtained before under
 the assumption that 
the distance inside the base graph can be ignored \cite{Harris}.
The authors argued that the approximation was a good one on the grounds that,
although the base graphs are critical,
 the average number of vertices in them is actually rather small.
 Our exact results show that their conclusion is in 
fact correct, and the distance in the base graphs does not affect the 
Hausdorff dimension.  The second case is $c=-2$ quantum gravity for which 
$\gamma_0=-1$ and there is overwhelming evidence that $d_{H_0}=(3+\sqrt{17})/2
\approx 3.562$ so that,
by Fisher scaling, $\eta_0\approx5.562 $. At the polymerized critical point we get
$\gamma=1/2$, $d_H=2$ and $\eta=1$ which is the same as in the branched 
polymer phase.

It is interesting that, at least in the case $\eta_0>3$, the known
exponents at the polymerized critical point coincide with those
of the continuously critical branched polymers studied in \cite{bialas,correia}.
It is easy to show, using the standard calculation of $\C H$ for
branched polymers that, provided only $\gamma>0$, a branched polymer
has $d_H=1/\gamma$, $\eta=1$, $d_h=2$, just as for the case  of
multi-critical branched polymers. The  multi-critical branched polymers
have $\gamma>1/2$ whereas the   continuously critical branched polymers
can have $\gamma <1/2$ which is what happens at the polymerized critical point. Of course we have not determined
$d_h$ or the spectral dimension $d_s$ for the polymerized critical point
but it is at least plausible that it does indeed fall in the same
universality class as the continuously critical branched polymers. This is precisely because of the argument given above; namely that the average number of vertices in a base graph is rather small.



\vspace{1 truecm}


\begin{thebibliography}{99}
\bibitem{das}S. R. Das, A. Dhar, A. M. Sengupta and S. R. Wadia,
Mod. Phys. Lett. A5 (1990) 1041.

\bibitem{gaume}L. Alvarez-Gaume, J. L. F. Barbon, and C. Crnkovic, Nucl. Phys.
B 394 (1993) 383.
\bibitem{korchemsky}G. P. Korchemsky, Phys. Lett. B 296 (1992) 323;
Mod. Phys. Lett. A7 (1992) 3081.

\bibitem{durhuus}B. Durhuus, Nucl. Phys. B 426 (1994) 203.

\bibitem{adj}J. Ambj\o rn, B. Durhuus and T. Jonsson, Mod. Phys. Lett. A9
(1994)
1221.
\bibitem{klebanov}I. R. Klebanov, Phys. Rev. D51 (1995) 1836;\\
I. R. Klebanov and A. Hashimoto, Nucl. Phys. B434 (1995) 264.

\bibitem{tjjfw} T. Jonsson and J.F.Wheater, Phys. Lett. B 345 (1995) 227.
\bibitem{Harris}M.G.Harris and J.Ambj\o rn, Nucl. Phys. B 474 [FS]
(1996) 575.

\bibitem{book}J. Ambj\o rn, B. Durhuus and T. Jonsson, \emph{Quantum Geometry},
Cambridge Monographs on Mathematical Physics, Cambridge 1997.
\bibitem{AmbA} J.Ambj\o rn and Y.Watabiki, Nucl. Phys. B445 (1995) 129.
\bibitem{AmbE} J.Ambj\o rn, J.Jurkiewicz and Y.Watabiki, Nucl. Phys. B454 
(1995) 313.

\bibitem{BP2}J.Ambj\o rn, B.Durhuus and T.Jonsson, Phys. Lett. B244
(1990) 403.
\bibitem{AmbC} J.Ambj\o rn et al, Nucl. Phys. B511 (1998) 673-710.

\bibitem{Kawai} H.Kawai, N.Kawamoto, T.Mogami and Y.Watabiki, 
Phys. Lett. B306 (1993) 19; Y.Watabiki, Nucl. Phys. B441 (1995) 119.

\bibitem{bialas}{P. Bialas and Z. Burda, Phys. Lett. B 384 (1996) 75.}

\bibitem{correia}J.D. Correia and J.F. Wheater, Phys. Lett. B422 (1998) 76.
%




 

\end{thebibliography}
\end{document}